\documentclass[12pt,psfig]{article}
\usepackage{amssymb}
\usepackage{amsmath}
\usepackage{epsfig}
\makeatletter
\def\@cite#1#2{{[{#1}]\if@tempswa\typeout
{IJCGA warning: optional citation argument
ignored: `#2'} \fi}}

\newcount\@tempcntc
\def\@citex[#1]#2{\if@filesw\immediate\write\@auxout{\string\citation{#2}}\fi
  \@tempcnta\z@\@tempcntb\m@ne\def\@citea{}\@cite{\@for\@citeb:=#2\do
            {\@ifundefined
       {b@\@citeb}{\@citeo\@tempcntb\m@ne\@citea\def\@citea{,}{\bf
?}\@warning
       {Citation `\@citeb' on page \thepage \space undefined}}%
    {\setbox\z@\hbox{\global\@tempcntc0\csname b@\@citeb\endcsname\relax}%
     \ifnum\@tempcntc=\z@ \@citeo\@tempcntb\m@ne
       \@citea\def\@citea{,}\hbox{\csname b@\@citeb\endcsname}%
     \else
      \advance\@tempcntb\@ne
      \ifnum\@tempcntb=\@tempcntc
      \else\advance\@tempcntb\m@ne\@citeo
      \@tempcnta\@tempcntc\@tempcntb\@tempcntc\fi\fi}}\@citeo}{#1}}
\def\@citeo{\ifnum\@tempcnta>\@tempcntb\else\@citea\def\@citea{,}%
  \ifnum\@tempcnta=\@tempcntb\the\@tempcnta\else
   {\advance\@tempcnta\@ne\ifnum\@tempcnta=\@tempcntb \else
\def\@citea{--}\fi
    \advance\@tempcnta\m@ne\the\@tempcnta\@citea\the\@tempcntb}\fi\fi}
\makeatother

\def\f#1#2{\frac{#1}{#2}}

\newcommand{\gsim}{\lower.7ex\hbox{$\;\stackrel{\textstyle>}{\sim}\;$}}
\newcommand{\lsim}{\lower.7ex\hbox{$\;\stackrel{\textstyle<}{\sim}\;$}}
\newcommand{\be}{\begin{equation}}
\newcommand{\ee}{\end{equation}}
\newcommand{\bea}{\begin{eqnarray}}
\newcommand{\eea}{\end{eqnarray}}

\usepackage{amssymb}

\def\baselinestretch{1}
\begin{document}
\catcode`@=11
\newtoks\@stequation
\def\subequations{\refstepcounter{equation}%
\edef\@savedequation{\the\c@equation}%
  \@stequation=\expandafter{\theequation}
  \edef\@savedtheequation{\the\@stequation}
  \edef\oldtheequation{\theequation}%
  \setcounter{equation}{0}%
  \def\theequation{\oldtheequation\alph{equation}}}
\def\endsubequations{\setcounter{equation}{\@savedequation}%
  \@stequation=\expandafter{\@savedtheequation}%
  \edef\theequation{\the\@stequation}\global\@ignoretrue

\noindent}
\catcode`@=12
\begin{titlepage}
\title{{\bf Modified gravity, Dark Energy and MOND}}
\vskip2in

\author{
{\bf Ignacio Navarro$$\footnote{\baselineskip=16pt E-mail: {\tt
i.navarro@damtp.cam.ac.uk}}} $\;\;$and$\;\;$ {\bf Karel Van
Acoleyen$$\footnote{\baselineskip=16pt E-mail: {\tt
karel.van-acoleyen@durham.ac.uk}}}
\hspace{3cm}\\
$$ {\small $^{*}$DAMTP, University of Cambridge, CB3 0WA Cambridge, UK}\\
{\small $^{\dagger}$IPPP, University of Durham, DH1 3LE Durham, UK}.
}

\date{}
\maketitle
\def\baselinestretch{1.15}

\begin{abstract}
\noindent

We propose a class of actions for the spacetime metric that
introduce corrections to the Einstein-Hilbert Lagrangian depending
on the logarithm of some curvature scalars. We show that for some
choices of these invariants the models are ghost free and modify
Newtonian gravity below a characteristic acceleration scale given
by $a_0 = c\mu$, where $c$ is the speed of light and $\mu$ is a
parameter of the model that also determines the late-time Hubble
constant: $H_0 \sim \mu$. In these models, besides the massless
spin two graviton, there is a scalar excitation of the spacetime
metric whose mass depends on the background curvature. This
dependence is such that this scalar, although almost massless in
vacuum, becomes massive and effectively decouples when one gets
close to any source and we recover an acceptable weak field limit
at short distances. There is also a  (classical) ``running'' of
Newton's constant with the distance to the sources and gravity is
easily enhanced at large distances by a large ratio. We comment on
the possibility of building a model with a MOND-like Newtonian
limit that could explain the rotation curves of galaxies without
introducing Dark Matter using this kind of actions. We also
explore briefly the characteristic gravitational phenomenology
that these models imply: besides a long distance modification of
gravity they also predict deviations from Newton's law at short
distances. This short distance scale depends on the local
background curvature of spacetime, and we find that for
experiments on the Earth surface it is of order $\sim 0.1mm$,
while this distance would be bigger in space where the local
curvature is significantly lower.

\end{abstract}

\vspace{3cm}

\vskip-24.5cm \rightline{} \rightline{DAMTP-2005-129}\rightline{DCPT/05/154}\rightline{IPPP/05/77}
\end{titlepage}
\setcounter{footnote}{0} \setcounter{page}{1}

\baselineskip=20pt

\section{Introduction}

The relative importance of the gravitational interaction increases
as we consider larger scales, and it is at the largest scales that
we can measure where the observed gravitational phenomena do not
agree with our expectations. The Hubble constant measuring the
rate of expansion of the Universe does not fall with time as
predicted by General Relativity (GR) for a Universe that contains
only known forms of matter, and the dynamics of galaxies seem to
require much more matter than observed if explained in terms of
GR. The most common approach to these problems is to assume the
presence of unseen forms of energy that bring into agreement the
observed phenomena with GR. The standard scenario to explain the
dynamics of galaxies consists in the introduction of an extra
weakly interacting massive particle, the so-called Cold Dark
Matter (CDM), that clusters at the scales of galaxies and provides
the required gravitational pull to hold them together. The
explanation of the observed expansion of the universe requires
however the introduction of a more exotic form of energy, not
associated with any form of matter but associated with the
existence of space-time itself: vacuum energy. And while CDM can
be regarded as a natural possibility given our knowledge of
elementary particle theory, the existence of a non-zero but very
small vacuum energy remains an unsolved puzzle for our high-energy
understanding of physics. However, the apparent naturalness of the
CDM hypothesis finds also problems when one descends to the
details of the observations. Increasingly precise simulations of
galaxy formation and evolution, although relatively successful in
broad terms, show well-known features that seem at odds with their
real counterparts, the most prominent of which might be the
``cuspy core'' problem and the over-abundance of substructure seen
in the simulations (see $e.g.$ \cite{Ostriker:2003qj}). But,
despite of this, the main problem that the CDM hypothesis faces is
probably to explain the correlations of the relative abundances of
dark and luminous matter that seem to hold in a very diverse set
of astrophysical objects \cite{McGaugh:2005er}. These correlations
are exemplified in the Tully-Fisher law \cite{Tully:1977fu} and
can be interpreted as pointing to an underlying acceleration
scale, below which the Newtonian potential changes and gravity
becomes stronger. This is the basic idea of MOND (MOdified
Newtonian Dynamics), a very successful phenomenological
modification of Newton's potential proposed in 1983
\cite{Milgrom:1983ca} whose predictions for the rotation curves of
spiral galaxies have been realised with increasing accuracy as the
quality of the data has improved \cite{Sanders:2002pf}.
Interestingly, the critical acceleration required by the data is
of order $a_0 \sim c H_0$ where $H_0$ is today's Hubble constant
and $c$ the speed of light (that we will set to 1 from now on).
The problem with this idea is that MOND is just a modification of
Newton's potential so it remains silent in any situation in which
relativistic effects are important. Efforts have been made to
obtain MONDian phenomenology in a relativistic generally covariant
theory by including other fields in the action with suitable
couplings to the spacetime metric \cite{Bekenstein:2004ne} (see also \cite{Drummond:2001rj} for other approaches to galactic dynamics without Dark Matter). But
these models do not address in a unified way the Dark Energy and
Dark Matter problems, while a common origin is suggested by the
observed coincidence between the critical acceleration scale and
the Dark Energy density. In this paper we will propose a class of
generally covariant actions, built only with the metric, that have
the right properties to address these problems in a unified way,
and where the relation $a_0 \sim H_0$ finds a natural explanation.
The theories we will consider modify gravity in the infrared,
making it stronger below a characteristic acceleration scale, but
this is not their only characteristic feature. When we are in a
situation in which the dominant gravitational field is external,
like in table-top experiments on Earth (that measure the
gravitational field of some probes embedded in the dominant
background gravitational field of the Earth), we can also expect
short distance modifications of Newton's potential.

Regarding the long distance modifications, the source-dependent
characteristic distance beyond which Newtonian gravity is modified in these
theories, that we shall call $r_c$, is given by
\be
\f{G_NM}{r_c^2} = \f{\mu}{2}\,, \label{MOND}
\ee
where $G_N$ is Newton's constant, $M$ the mass of the source and $\mu$ is a
parameter of the model that also
determines the late-time Hubble constant: $H_0 \sim \mu$. This makes
these models promising candidates to build a theory with a MOND-like
Newtonian limit that could address the dynamics of galaxies without
the need for Dark Matter. But when measuring
the
gravitational attraction between two probes in the external
dominant gravitational field of a massive
object of mass $M$, at a distance $r_d$ from its centre, we can also expect
short distance modifications of Newton's law for distances smaller than
\be
r_{SD}\sim \f{\mu r_d^3}{G_N M}.
\ee
If we plug in this expression the radius and mass of the Earth (with
$\mu \sim H_0$), we get
that for table top tests of Newton's law performed on the Earth
surface $r_{SD} \sim 10^{-2}cm$. This range is very interesting
because it is the range currently being probed by experiments
\cite{Adelberger:2003zx}. But notice that the phenomenology of these theories is very
different from the one expected from other
theoretical considerations that also suggest a deviation from Newton's
law at that scale motivated by the cosmological constant
problem\footnote{It is well known that naturalness arguments lead to the
expectation that new
  physics associated with electro-weak symmetry breaking, besides the
  Higgs boson, should be seen in the LHC. Otherwise the
  electroweak scale becomes
unstable under
  quantum corrections. Applying the same logic to the gravitational
  sector one would expect
  new gravitational phenomena to kick in at the vacuum energy scale that
would
  cut-off the quantum divergences contributing to the vacuum
  energy. This hypothetical new physics should
  be seen in sub-mm measurements of Newton's potential \cite{Beane:1997it}.} or the gauge hierarchy problem \cite{Arkani-Hamed:1998rs}. The fact
that this scale is the same in both cases is just a numerical
coincidence. If we performed the same experiment on space, in the
neighbourhood of the Earth's orbit for instance where the dominant
gravitational field is that of the Sun, the relevant mass and
distance we should use in the previous estimation is the Sun's mass
and the Sun-Earth distance. In this case the ``short distance''
corrections would be expected in our theory at distances less than
about $10^4 m$! This however does not mean that there should be
big modifications to the motion of the planets or other celestial
bodies. When the gravitational field we are measuring is that of
the Sun, the corrections are suppressed at distances less than
$r_c$ that is in this case of the order of $10^3 AU \sim
10^{11}km$.

These characteristic experimental signatures arise in our
theory because of the presence of an extra scalar excitation of
the spacetime metric besides the massless spin two graviton. But
while the graviton remains massless, the extra scalar has a mass
that depends on the background curvature. This dependence is such
that for the models that we will be interested on, those that
modify gravity at large distances, this field becomes massive and
effectively decouples when the background curvature is large, and
in particular when we approach any source.

In the next section we
will briefly review the results we obtained in
\cite{Navarro:2005gh,Navarro:2005da} studying models that involve
inverse powers of the curvature in the action,
giving some general expressions and discussing the generic
features of the framework we will use for modifying gravity in the
infrared. In particular we will focus on a class of models that had been
proposed to address the acceleration of the Universe \cite{Carroll:2004de} and
also modify gravity at large distances \cite{Navarro:2005gh,Navarro:2005da}. This discussion
will enable us to motivate the class of actions that we will present in the
third section, that depend on the logarithm of some curvature invariants. We will see that the theories that we propose in this
section modify gravity at the MOND characteristic acceleration
scale, and the gravitational interaction can easily become stronger at
large distances. We will explore briefly the characteristic
gravitational phenomenology expected in these models, discussing
possible tests of these theories. In the fourth
section we offer the conclusions. We will comment on
further generalisations of the proposed actions and on the generic
phenomenological features expected in the class of models that
modify gravity at the MOND acceleration scale. We will also comment on the possibility
of obtaining these theories as an effective action for the spacetime
metric that takes into account strong renormalisation effects in the
infrared that might appear in GR.

\section{Modified gravity as an alternative to Dark Energy}

Recently, models involving inverse powers of the curvature have
been proposed as an alternative to Dark Energy
\cite{Carroll:2004de,Capozziello:2003tk}. In these models one
generically has more propagating degrees of freedom in the
gravitational sector than the two contained in the massless
graviton in GR. The simplest models of this kind add inverse
powers of the scalar curvature to the action ($\Delta {\cal
L}\propto 1/R^n$), thereby introducing a new scalar excitation in
the spectrum. For the values of the parameters required to explain
the acceleration of the Universe this scalar field is almost
massless in vacuum and one might worry about the presence of a new
force contradicting Solar System experiments. However in a recent
publication \cite{Navarro:2005gh} we showed that models that
involve inverse powers of other invariants, in particular those
that diverge for $r\rightarrow 0$ in the Schwarzschild solution,
generically recover an acceptable weak field limit at short
distances from sources by means of a $screening$ or $shielding$ of
the extra degrees of freedom at short distances
\cite{Navarro:2005da}.

But let us start by discussing the linearisation and vacuum
excitations obtained from generic actions built with the Ricci
scalar and the scalars \be P \equiv R_{\mu\nu}R^{\mu\nu}\,\;\;
{\rm and}\;\;\; Q \equiv
R_{\mu\nu\lambda\rho}R^{\mu\nu\lambda\rho}.\ee If the Lagrangian
is a generic function ${\cal L}=F(R,P,Q)$ the equations of motion
for the metric will be of fourth order and we can expect that the
particle content of the theory will have eight degrees of freedom:
two for the massless graviton, one in a scalar excitation and five
in a ghost-like massive spin two field \cite{Hindawi:1995cu}.
Expanding the action in powers of the curvature perturbations it
can be seen that at the bilinear level the linearisation of the
theory over a maximally symmetric spacetime will be the same as
that obtained from \cite{Hindawi:1995cu,Chiba:2005nz} \bea S=\int
\!\!d^4x\sqrt{-g}\frac{1}{16\pi
  G_N}\left[-\Lambda + \delta R
+\frac{1}{6m_0^2}R^2-\frac{1}{2m_2^2}C^{\mu\nu\lambda\sigma}C_{\mu\nu\lambda\sigma}\right]
\label{expand-action},\eea where $C_{\mu\nu\lambda\sigma}$ is the
Weyl tensor and we have defined \bea \Lambda &\equiv& \left<F-RF_R
+R^2\left(F_{RR}/2-F_P/4-F_Q/6\right)+R^3\left(F_{RP}/2+F_{RQ}/3\right)
\right.\nonumber \\
&&\left.+R^4\left(F_{PP}/8+F_{QQ}/18+F_{PQ}/6\right)\right>_0 \\
\delta &\equiv&
\left<F_R-RF_{RR}-R^2\left(F_{RP}+2F_{RQ}/3\right)
\right.\nonumber \\
& & \left.
-R^3\left(F_{PP}/4+F_{QQ}/9+F_{PQ}/3\right)\right>_0 \label{delta}\\
m_0^{-2}& \equiv&
\left<\left(3F_{RR}+2F_P+2F_Q\right)+R\left(3F_{RP}+2F_{RQ}\right)\right.\nonumber
\\
& & \left.
+R^2\left(3F_{PP}/4+F_{QQ}/3+F_{PQ}\right)\right>_0\label{m0}\\
m_2^{-2} &\equiv & -\left<F_P+4F_Q\right>_0. \eea Here $<...>_0$
denotes the value of the corresponding quantity on the background
and $F_R\equiv \partial_RF$, etc... One can see that the situation
for the perturbations over vacuum in any modified theory of
gravity (built with $R$, $P$ and $Q$) will be the same as in Einstein gravity supplemented with
curvature squared terms. It is well known that for the action
(\ref{expand-action}) the mass of the ghost is $\sim m_2$ and that
of the scalar is $\sim m_0$. So in the case in which
$F(R,P,Q)=F(R,Q-4P)$, $m_2^{-2}=0$ and there is no ghost in the
spectrum, but there is still the extra scalar. It is easy to check
that in the models that explain the acceleration of the Universe
by using actions that involve inverse powers of the scalar
curvature alone, the mass of the scalar is proportional to some
positive power of the scalar curvature \cite{Chiba:2003ir}. And notice
that including other terms in $F$ and fine-tuning the parameters
we can make $m_0^{-2}=0$ in vacuum, as suggested in
\cite{Nojiri:2003ft}, by including $R^2$ corrections to the action
besides the $1/R$ or ${\rm Log}(R)$ ones. But the infinite mass
(or absence) of the scalar is a property only of a particular
background in these models. If we evaluate the expression for
$m_0$ for $F=R-\mu^4/R + R^2/m_s^2$ for instance, we get \be
m_0^{-2} =
6\left(-\f{\mu^4}{\left<R^3\right>_0}+\f{1}{m_s^2}\right), \ee and
choosing a particular value of $m_s$, namely $m_s^2= 3\sqrt{3}
\mu^2$, we can make $m_0\rightarrow \infty$ in vacuum, where
$\left<R\right>_0 =\sqrt{3} \mu^2$. But we see that when the
scalar curvature is bigger than its vacuum value, the scalar mass
returns to its natural value, $m_0 \sim m_s \sim \mu$, which shows
that in these models the scalar is still present in general, and
its mass is very small in most situations\footnote{This is in
contrast with the absence of the ghost for actions in which
$F(R,P,Q)=F(R,Q-4P)$. In this case $m_2^{-2}=0$ independently of
the background curvature.}. But for the models that include only
inverse powers of the curvature, besides the Einstein-Hilbert
term, it is however possible that in regions where the curvature
is large the scalar has naturally a large mass and this could make
the dynamics to be similar to those of GR \cite{Cembranos:2005fi}.
But the scalar curvature, although bigger than its mean
cosmological value, is still very small in the Solar System for
instance. So, although a rigorous quantitative analysis of the
predictions of these models for observations at the Solar System
level is still lacking in the literature, it is not clear that
these models constitute a viable alternative to Dark Energy
because one can expect that the effects of this extra field should
have been observed\footnote{There is an alternative formulation of these models in which the connection and the metric are varied independently, the so-called Palatini formalism, and in this case these theories might be viable \cite{Vollick:2003aw}.}.

But the story is different if we include inverse powers of curvature
invariants, like $Q$, that grow at short distances in the Schwarzschild
solution. In this case the mass of the extra scalar field is guaranteed to grow as we
approach any source and we recover Einstein gravity at short distances
\cite{Navarro:2005gh,Navarro:2005da}. For studying these effects we focused on actions of the type
\be S=\int \!\!d^4x\sqrt{-g}\frac{1}{16\pi
  G_N}\left[R-\frac{\mu^{4n+2}}{(aR^2-4P+Q)^n}\right]\,.\label{action1}\ee
These actions, with $\mu$ of the order of the late time Hubble
constant $H_0$, were proposed in \cite{Carroll:2004de} motivated
by the Dark Energy problem and have cosmological solutions
providing a good fit to the SN data \cite{Mena:2005ta}. When
linearising over vacuum we find, besides the usual massless
graviton, a scalar excitation of the metric with a mass of order
$m_0 \sim \mu \sim H_0$. But in the expanded action higher order
non-renormalisable operators appear suppressed by inverse powers
of the background curvature. This means that the linearisation
will break down when the energy of the fluctuations is extremely
small: the strong coupling scale in this theory is $\Lambda_s\sim
\left(M_p H_0^{2n+4}/\mu^{2n+1}\right)^{1/4} \sim
\left(M_pH_0^3\right)^{1/4}$ \cite{Navarro:2005da}. This, in turn,
means that for a spherically symmetric solution, the linearisation
over vacuum will break down at a huge distance from any source, at
the so-called Vainshtein radius, that is for these theories $r_V
\sim \left(G_NM \mu^{2n+1}/H_0^{2n+4}\right)\sim
(G_NM/H_0^3)^{1/4}$ \cite{Navarro:2005da}. At smaller distances we
can not trust the results obtained using this expansion. It is
also important to keep in mind that in this long-distance regime,
where we can apply the linearisation of the action over vacuum,
the effective Planck mass is given by (see \cite{Navarro:2005da})
\be
M_{p(eff)}^2=M_p^2\left(\delta+\f{\left<R\right>_0}{3m_0^2}\right)=\left(1+\f{6n(a-1)}{(n+1)(6a-5)}\right)M_p^2
\label{Mpeff} \ee where $M_p^2 = (8\pi G_N)^{-1}$.

As we have said, at distances less than $r_V$ we enter a
non-perturbative regime where we can no longer use the linearised
action over vacuum because higher order non-renormalisable
operators become more important than those involving only two
powers of the fluctuations. But we can get some information about
the short distance behaviour of the solutions by noticing that the
extra term that the modification introduces will be unimportant at
short distances in the spacetime of a spherically symmetric mass.
The reason is that this term will always be suppressed by inverse
powers of $Q$, that for the Schwarzschild solution reads
$Q=48(G_NM)^2/r^6$ and grows at short distances. This tells us
that any spherically symmetric solution for which the curvature grows at short radius will
converge to the Schwarzschild one with a Planck mass given simply
by $M_p^2 = (8\pi G_N)^{-1}$. We can then make a different kind of
expansion, a weak field expansion over the Schwarzschild solution
that shows that in these models Newtonian gravity is modified
beyond a distance given by \cite{Navarro:2005gh} \be r_c^{3n+2}
\equiv \frac{\left(G_N M\right)^{n+1}}{\mu^{2n+1}}. \ee In fact,
for these theories, the spherically symmetric solution, in an
expansion in powers of $r/r_c$ reads: \bea ds^2&\simeq
&-\left[1-\frac{2G_NM}{r}\left( 1 - \alpha
\left(\frac{r}{r_c}\right)^{6n+4}+ {\cal
O}\left((r/r_c)^{12n+8}\right)\right)\right]dt^2
\\&+&\left[1-\frac{2G_NM}{r}\left( 1 + \frac{\alpha(6n+3)}{2}
\left(\frac{r}{r_c}\right)^{6n+4}+ {\cal
O}\left((r/r_c)^{12n+8}\right)\right)\right]^{-1}dr^2+r^2d\Omega^2_2\,,\nonumber\eea
where $M$ is the mass of the object sourcing the field and
$\alpha\equiv \frac{n(1+n)}{(6n+3)2^{4n}3^n}$. 
\begin{figure}[h]
  \begin{center}
    \epsfig{file=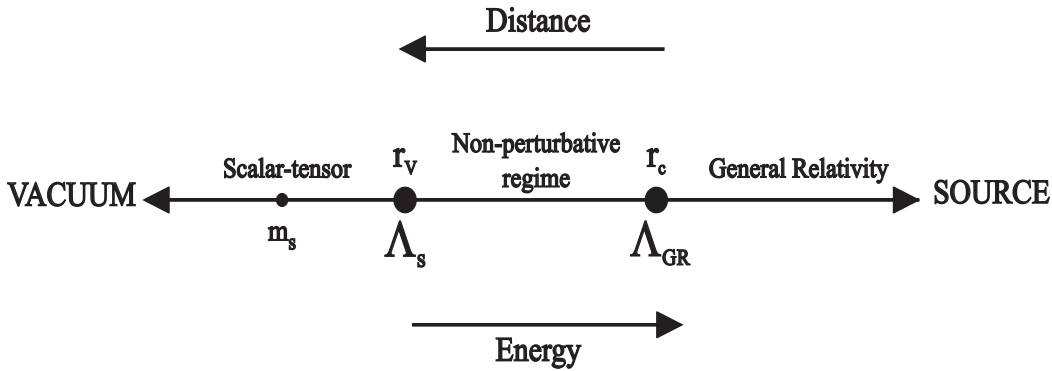,height=5cm,width=14cm}
\caption{Schematic representation of the structure of the theory
on the energy/distance scale.}
\end{center}
\end{figure}

The distance $r_c$
is very large, at least of the order of parsecs for a star like
the Sun (assuming $n\geq 1$), so from this expansion we see that
the corrections that we can expect in these theories at the Solar
System level are very small. Unfortunately, this short distance
expansion breaks down for distances of order $r_c$, and notice
that $r_c \ll r_V$. So in these theories we are led to the
following picture: at small energies ($E < \Lambda_s$) or large
distances ($r> r_V$) we have a scalar tensor theory, with an
almost massless scalar and an effective Planck mass given by
(\ref{Mpeff}).
As we increase the energy of the fluctuations or go to shorter
distances we enter a non-perturbative phase, but at even higher
energies ($E>\Lambda_{GR}$) or smaller distances ($r<r_c$) we can
neglect the effects of the modification and we recover the
standard GR dynamics, with a Planck mass given by $M_p$. Here $\Lambda_
{GR}=\left(M_p^{n+1}\mu^{2n+1}\right)^{1/(3n+2)}$ is the energy
scale associated with this high energy recovery of Einstein
gravity. This situation is depicted in fig.1. Since all the
expansions that we have used break down in this non-perturbative
intermediate regime, at this point we can only assume that the
dynamics are consistent for this range of energies/distances, and
that one can obtain a consistent matching between the long
distance and short distance solutions. However we regard this as a
model dependent issue, and we can extract already a lot of useful
information just knowing the dynamics in the low and high energy
regimes.

In terms of the particle content of the linearised theory we can
get some ``non-perturbative'' insight in the reasons behind the
behaviour of these solutions by evaluating the expression for
$m_0$ in them. We see then that we can expect that the mass of the
extra degree of freedom that the modification introduces has a
contribution in the spacetime of a spherically symmetric mass that
goes like \cite{Navarro:2005da} \be \delta_{Source} m_s(r) \sim
\f{Q^{\f{n+1}{2}}}{\mu^{2n+1}}  \sim
\f{(G_NM)^{n+1}}{\mu^{2n+1}r^{3n+3}}. \ee This effective mass
decouples the scalar at short distances, where $Q\gg \mu^4 $, and
we recover Einstein gravity in this domain. It is also worth to
point out that one recovers the distance $r_c$ as the distance at
which $m_s(r)\sim r^{-1}$, with $m_s(r)$ given by the expression
above. Since $m_s(r)$ grows faster than $r^{-1}$, at $smaller$ distances the scalar effectively decouples.
But the $r$-dependent mass of the scalar field is not the only
effect of the modification: as we have seen at large distances
from sources the effective Planck mass is given by
eq.(\ref{Mpeff}) while at short distances it is simply $M_p$. So
there is a rescaling or ``running'' of the Planck mass with the
distance to the sources.

However, one can see that these theories, although capable of
explaining the acceleration of the Universe without Dark Energy,
are not capable of explaining the dynamics of galaxies without
Dark Matter. First, the critical distance at which the
modification becomes important ($r_c$)  is too large, since the
data consistently indicate that any such theory should have
noticeable effects at a distance $r_c$  given by eq.(\ref{MOND}).
And second, the effective Planck mass that we obtain at large
distances is not significantly reduced with respect to the one
that we get at short distances except for a very small range of
values of $a$ that, at least for the $n=1$ case, are not
phenomenologically acceptable \cite{Mena:2005ta}. We need the
reduction of the effective Planck mass in vacuum in order to get a
large enhancement of the gravitational interaction at large
distances as required to have any hope to fit the data without
introducing Dark Matter. The critical distance that we obtain in
these theories does however suggest that in the $n\rightarrow 0$
limit the modification becomes important at a distance that
corresponds to the MOND characteristic acceleration. This
motivates the use of Logarithmic actions as a possibility for
obtaining a modification of gravity as an alternative to Dark
Matter. A first assessment of this possibility is the goal of the
next section.

\section{Logarithmic actions and modified gravity as an alternative to
  Dark Matter}

As we said, the considerations of the previous section motivate the
introduction of an action depending on the logarithm of the curvature
in order to get a modification of gravity at the MOND characteristic
acceleration scale. In this section we will discuss what appears to be
the simplest possibility, namely actions of the type
\be
S=\int \!\!d^4x\sqrt{-g}\frac{1}{16\pi
  G_N}\left\{R-\mu^{2}{\rm
Log}\left[f(R,Q-4P)\right]\right\}\,,\label{actionMOND} \ee where
for the function $f$ we will only assume that \be f\rightarrow
0\;\;\; {\rm for} \;\;\; R_{\mu\nu\lambda}^{\sigma}\rightarrow
0\,,\ee and we can approximate \be f\simeq Q/Q_0\;\;\; {\rm when} \;\;\; Q\gg R^2,P.\ee
In this case Minkowski spacetime will not be a solution of the
theory but there will typically exist de Sitter solutions with
$H_0 \sim \mu$. Notice that the addition of a cosmological constant will not change the form of the action\footnote{Adding a cosmological constant would simply redefine the function $f$. When writing the action in this form we are of course assuming that the mass scales appearing in $f$ are ``reasonable'' in terms of $\mu$ so that $H_0\sim \mu$.}. As we will see we can apply the discussion of the
previous section to these theories simply taking $n=0$ in the
relevant formulae. We will study in the next subsection the
behaviour of the spherically symmetric solutions of this theory at
short distances from sources, and we will show that in this domain
the corrections to the Schwarzschild geometry are small. Also, it
will be made clear that the Newtonian potential has large
corrections at a distance given by $r_c$ in eq.(\ref{MOND}). In
subsection 3.2 we will study the linearisation of this theory in
vacuum. We will give the conditions for the stability of de Sitter
space and we will see that the effective Planck mass in vacuum can
easily be reduced with respect to the one at short distances by a
large ratio, and in this case gravity becomes significantly
stronger at large distances. The expansion of the action in vacuum
is the relevant one to apply in some dynamical situations at very
large distances from sources and in late-time cosmology, but one
should keep in mind that this linearisation breaks down when one
approaches any source or when the ambient curvature is large
(like, for instance, in the early Universe). In subsection 3.3 we
will discuss some of the characteristic experimental signatures
that these theories imply.

\subsection{Short distance solution}

In this subsection we will follow the same strategy as in
\cite{Navarro:2005gh} in order to study the behaviour of the
solutions at short distances from sources. The spherically
symmetric solutions for this theory  are obtained by solving the
equations \be G_{\mu\nu}+\mu^{2}H_{\mu\nu}=0\,,\ee where
$G_{\mu\nu}$ is the usual Einstein tensor and $\mu^{2}H_{\mu\nu}$
is the extra term generated by the logarithmic part of the action.
One can see that this extra term, when evaluated in the
Schwarzschild solution, is subleading with respect to the terms
that appear in the Einstein tensor when $r\ll r_c$. This indicates
that as $r\rightarrow 0$ the corrections with respect to the
Schwarzschild geometry will be small. We can then consider a small
perturbation of the black hole geometry and solve at first order
in the perturbations. So we take the ansatze \be
ds^2=-\left[1-\frac{2G_NM}{r}+\epsilon
A(r)\right]dt^2+\left[1-\frac{2G_NM}{r}+\epsilon
B(r)\right]^{-1}dr^2+r^2d\Omega^2_2\,,\label{sdsol}\ee and treat
$\epsilon$ as a small expansion parameter. We can expand the full
equations in powers of $\epsilon$: \bea G_{\mu\nu}&=&
G_{\mu\nu}^{(0)}+\epsilon G_{\mu\nu}^{(1)}+\epsilon^2
G_{\mu\nu}^{(2)}+\ldots\,, \nonumber\,\\ H_{\mu\nu}&=&
H_{\mu\nu}^{(0)}+\epsilon H_{\mu\nu}^{(1)}+\epsilon^2
H_{\mu\nu}^{(2)}+\ldots\,.\eea Since the Schwarzschild solution
solves the ordinary Einstein equations, we have
$G_{\mu\nu}^{(0)}=0$. Treating $\mu^{2}$ as an order $\epsilon$
parameter, at first order in our expansion the equations for $A$
and $B$ become (from now on we set $\epsilon=1$) \be
G_{\mu}^{\nu(1)}=-\mu^{2}H_{\mu}^{\nu(0)}\,. \ee For the $tt$
component of this equation we find: \be
\frac{B+rB'}{r^2}=-\frac{\mu^2}{2G_NM}\left(13G_NM-8r+G_NM{\rm
  Log}\left[\frac{48 (G_NM)^2}{r^6Q_0}\right]\right)
\ee while the $rr$ component reads: \be
\frac{(2G_NMA-rB)(2G_NM-r)^{-1}+rA'}{r^2}=\frac{\mu^2}{2G_NM}\left(7G_NM-2r-G_NM{\rm
  Log}\left[\frac{48 (G_NM)^2}{r^6Q_0}\right]\right).
\ee We can solve the previous equations yielding \bea B(r)&
=&\left(\frac{2G_NM}{r}\right)\left(\frac{r}{r_c}\right)^4\left[2-\f{G_NM}{3r}\left(15+{\rm
  Log}\left[\frac{48(G_NM)^2}{Q_0r^6}\right]\right)\right],\nonumber \\
A(r)&
=&-\left(\frac{2G_NM}{r}\right)\left(\frac{r}{r_c}\right)^4\left[\f{4}{3}-\f{G_NM}{3r}\left(5-{\rm
  Log}\left[\frac{48(G_NM)^2}{Q_0r^6}\right]\right)\right],
\label{sdsolII}\eea where $r_c$ is now given by eq.(\ref{MOND}). So we
see that at short distances the corrections to Newton's potential
are indeed suppressed by powers of $(r/r_c)^4$. In fact, we can
consider our approximate solution as the first order in the Taylor
expansion in powers of $r/r_c$ of the exact solution, where the
next order in the expansion would be ${\cal
  O}\left((r/r_c)^8\right)$. At distances of order $r_c$ the expansion
breaks down, and it is clear that we can expect a significant
modification of Newton's potential for larger distances (or
smaller accelerations). A difference with respect to the actions
involving inverse powers of the curvature is that now the scalar
curvature does not go to zero at short distances. This is so
because the logarithmic part of the action does not go to zero for
$r\rightarrow 0$, as happened in the case of the inverse powers in
the previous section, but the scalar curvature (or the extra term
in the action) is of the order of the one in
vacuum so the effect is very small.

In the next subsection, we study the linearisation of this theory in
vacuum, which will enable us to obtain the ultra large distance
behaviour of the spherically symmetric solution corresponding to a
mass source.

\subsection{Linearisation in vacuum}

In vacuum, without any matter source, the action has de Sitter
solutions with curvature $R=12H_0^2$ where $H_0$ can be found as a
solution of \be 6H_0^2 = \mu^2 \left<{\rm
Log}f-6H_0^2\f{f_R-20H_0^2f_Q}{f}\right>_0. \label{hubble}\ee To
check the stability of these solutions one can consider the
modified Friedmann equation in vacuum obtained in this theory:
$\ddot{H}=h(H,\dot{H})$ and expand for small perturbations
$\dot{H}$ and $\tilde{H}\equiv H-H_0\,\,$ of the equation of
motion. We find: \be \ddot{H}\approx
h_H(H_0,0)\tilde{H}+h_{\dot{H}}(H_0,0)\dot{H}
=16H_0^2\left(\f{1}{4}+\f{C_1}{C_2}\right)\tilde{H}-3H_0\dot{H}\,,\ee
where we have defined \be C_1 \equiv \delta
+\frac{4H_0^2}{m_0^2}\;\;\; {\rm and} \;\;\; C_2 \equiv
-\frac{16H_0^2}{m_0^2}\,, \label{C1C2}\ee $m_0$ and $\delta$ have
been defined in the second section. One can easily check that the
fixed point ($\tilde{H}=0$, $\dot{H}=0$) of this dynamical system
is an attractor (repeller) if the coefficient in front of
$\tilde{H}$ is negative (positive), so de Sitter space will be
stable as long as \be 1+\f{4C_1}{C_2}< 0. \ee One would obtain the
same result by analysing the equation of motion of the propagating
modes, and it is instructive to do so. Because the action is a
function of $Q$ and $P$ only through the combination $Q-4P$, the
ghost is absent and there is just a
massive scalar field in the spectrum besides the massless graviton. So this field is the only
possible source of instability. It was shown in
\cite{Navarro:2005da} that in this case de Sitter spacetime is
stable as long as $m_s^2>-9H_0^2/4$, and the mass of the scalar
is \be m_s^2\equiv
-H_0^2\left(\f{25}{4}+16\f{C_1}{C_2}\right),\label{ms} \ee which
is consistent with the phase space analysis presented here and
also with the results of \cite{Faraoni:2005vk}, for $F(R)$
theories. But even when de Sitter space is unstable one should not
disregard the model. It has been shown that for the actions
involving inverse powers of the curvature de Sitter space is
unstable in many cases but the late time background corresponds to
a power law FRW cosmology that is nevertheless phenomenologically
interesting \cite{Carroll:2004de,Mena:2005ta}.

An important feature of this linearisation is that the effective
Planck mass that we obtain in vacuum controlling the coupling of
the spin two graviton, is now given by \be M_{p(eff)}^2=C_1 M_p^2
= \left<1-\mu^2\f{f_R-2Rf_Q}{f}\right>_0M_p^2.\label{Geff}\ee
 At short distances however the value of the Planck
mass is just $M_p$, as we saw using the short distance expansion over Schwarzschild geometry. But in vacuum $R^2\sim P \sim Q \sim \mu^4 \sim
H_0^4$ and we can expect that $C_1$ will depart significantly from
1. This means that when this linearisation is applicable, we get a
rescaled Planck mass with respect to the one we would infer at
short distances and one can expect an enhancement or suppression
of the gravitational interaction at large distances. In fact,
applying this linearisation, we get for a spherically symmetric
mass the solution \cite{Navarro:2005da}: \be ds^2 \simeq
-\left(1-\frac{8G^{(eff)}_NM}{3r}-H_0^2r^2\right)dt^2
+\left(1-\frac{4G^{(eff)}_NM}{3r}-H_0^2r^2\right)^{-1}dr^2+r^2d\Omega^2_2
\label{ldsol}\ee where $G^{(eff)}_N = G_N/C_1$ is the effective
long-distance Newton's constant. Remember that the
linearisation we are using to get this result breaks down at a
huge distance from any source, at the Vainshtein radius $r_V
\simeq (G_NM/H_0^3)^{1/4}$. At shorter distances we can not use
the linearised version of the theory, and the solution above is
only a good approximation for $r> r_V$. But we have seen that at
even shorter distances, $r< r_c$, we can use a different
expansion: the one that we considered in the previous sub-section
that shows that in this regime the solution,
eqs.(\ref{sdsol},\ref{sdsolII}), approaches the Schwarzschild one
with a Planck mass given by $M_p^2$. So also in these theories
there is a non-perturbative regime in the range of distances  $r_c
< r < r_V$, where one should get an interpolation between the
solutions eqs.(\ref{sdsol},\ref{sdsolII}) and eq.(\ref{ldsol}). In
this intermediate range of distances both the linearisation of the
theory over the vacuum and the short distance expansion over the
Schwarzschild solution break down, but we will assume that the
function $f(R,Q-4P)$ is such that a consistent matching exists.
When $C_1<1$ and Newton's constant is enhanced at large distances,
the interpolating regime would play the role of a ``dark halo''
surrounding any source.

Even though we have not solved the gravitational equations in this
non-perturbative regime, we would like to stress some important
characteristics of this intermediate range of distances that make
these theories promising candidates to build a MOND-like
modification of gravity that could explain the dynamics of
galaxies without recourse to Dark Matter. First, the modification
becomes important below a characteristic Newtonian acceleration
scale determined by a parameter ($\mu$) that also determines the
late-time Hubble constant, as consistently indicated by the data
\cite{Sanders:2002pf}. And second, the gravitational interaction
can be enhanced at long distances by a ratio that depends on the
particular function $f$ that we choose, but that can easily be of
the required magnitude. As an example we plot in fig.2 the ratio
$M^2_p/M_{p(eff)}^2 = 1/C_1$, for $f=(2R^2-4P+Q)/Q_0$, as a
function of ${\rm Log}[\mu^4/Q_0]/2$. The apparent mass
discrepancy that one would infer cosmologically or at large
distances is of the order of this ratio. There are further
corrections because the theory at large distances is of the
scalar-tensor type, with an almost massless scalar, so we get an
extra 4/3 factor in the Newtonian potential and the deflection of
light would also get an analogous correction factor.
\begin{figure}[h]
  \begin{center}
    \epsfig{file=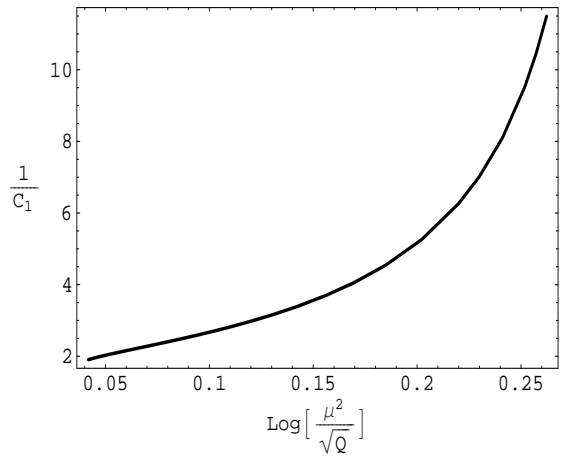,height=6cm,width=8cm}
\caption{Ratio of the effective Planck mass at short distances
($r<r_c$) over
  the one at large distances ($r>r_V)$ for a spherical mass in
the model
  (\ref{actionMOND}) with $f=(2R^2-4P+Q)/Q_0$, as a function of ${\rm Log}[\mu^2/Q_0^{1/2}]$. This represents roughly the enhancement
  of the gravitational interaction at large distances.}
\end{center}
\end{figure}

These theories make a diverse variety of characteristic
experimental predictions that can be used to test or falsify them.
In this section we have seen the approximate form of the solution
for a spherically symmetric mass at short distances ($r<r_c$) and
at long distances ($r>r_V$) in de Sitter space. Using these
approximate solutions and the linearised version of the theory,
once we have chosen a particular function $f$, we can already
compare the predictions of the theory with precision Solar System
measurements and a wealth of cosmological, astrophysical and large
scale structure data. But these are not the only testing grounds
for the theory. It is also possible to obtain predictions that
would differ from those of GR for laboratory experiments measuring
the gravitational interaction of small bodies. The discussion of
these experimental implications is the subject of the following
subsection.

\subsection{Experimental implications}

The theories we are studying in this section offer a diverse range
of characteristic experimental predictions differing from those of
GR that would allow their falsification. The most obvious tests
would come from the comparison of the predictions of the theory to
astrophysical and cosmological observations where the dynamics are
dominated by very small gravitational fields. But we would have
also some small effects for the motion of the planets or other
celestial bodies in the Solar System and as we have said short
distance modifications of Newton's law. In the following we will
briefly discuss separately these possible tests, and we will give
order-of-magnitude estimations of the expected effects.

\subsubsection{Short distance deviations from Newton's law}

Probably the most characteristic experimental signature of our
theory corresponds to deviations from Newton's law that should be
seen at short distances, where the precise distance at which the
modification is noticeable depends on the local background
curvature. While a detailed computation of these corrections will
be deferred to a future publication \cite{toappear}, to see why
this is so we can use the following arguments. As we have said,
for theories of the type (\ref{actionMOND}) there is an extra
scalar excitation of the spacetime metric besides the massless
graviton. The mass of this field is given in de Sitter space by
eq.(\ref{ms}), but it picks up an $r$-dependent contribution in
the spacetime of a spherically symmetric mass. In a generic
background, if we study a system in a scale that reduces the
space-time region of interest to be small enough, we can consider
an effective theory in which the extra degree of freedom has a
mass given by its local value. Now, if we evaluate the expression
for $m_0$ in the spacetime of a spherically symmetric mass we get
that at a distance $r_{d}$ of a massive object of mass $M$, the
mass of the scalar is locally of order \be m_s^2 \sim \f{Q}{\mu^2}
\sim \f{(G_NM)^2}{r_{d}^6\mu^2},\label{msrsd} \ee where we are
assuming that $Q\gg R^2,P$ so $f\simeq Q/Q_0$. So in the effective
gravitational theory that we should apply on the Earth surface
there is, besides the massless spin two graviton, an extra scalar
field with gravitational couplings and a mass given roughly by the
expression above. A peculiar feature of this local effective
theory on a Schwarzschild background is that there will be a
preferred direction and this will be reflected in an anisotropy of
the force that this scalar excitation will mediate
\cite{toappear}. But for the purposes of this section, an
estimation of the expected order-of-magnitude of the corrections,
we will simply focus on the value of the mass of the scalar.
From effective field theory arguments we can then expect short distance modifications of Newton's law,
suppressed by $r_{SD} \sim m_s^{-1} \sim {\cal O}(0.1\; mm)$, when
measuring the gravitational field of some probes on the Earth
surface. This order of
magnitude is very interesting because it is the one currently
being probed by experiments \cite{Adelberger:2003zx}. As we
said there are also other motivations for expecting short distance modifications of gravity at this scale \cite{Beane:1997it,Arkani-Hamed:1998rs}, but as we also said the coincidence of these scales is just
a coincidence. If one was to measure the gravitational field of a
small object in space, in the neighbourhood of the Earth's orbit
but far from the Earth, the relevant mass and distance we should
apply in eq.(\ref{msrsd}) is the mass of the Sun and the Sun-Earth
distance, since the Sun provides the dominant gravitational field
in that situation. In this case we get $m_s^{-1} \sim 10^4\;m$.
But notice that to observe a significant modification in the
gravitational field of an object we have to measure it at a
distance bigger than $r_c$ for that object, otherwise the
self-shielding of the extra scalar excitation induced by the
object itself is enough to switch off the modification. This means
that locally in the inner Solar System we could only see significant modifications in the
gravitational field of objects whose characteristic distance $r_c$
is smaller than $10^4\; m$, so its mass would have to be below
$\sim 10^{9}\; kg$. As an example we can mention that for an
object of mass $10^3\; kg$ orbiting the Sun at the same distance
as the Earth, one can expect modifications of its gravitational
field starting at a distance $r_c \sim 10\; m$ (at $shorter$
distances the scalar effectively decouples because of the
gravitational field of the object itself) and extending to
$r_{SD}\sim 10^4 \; m$ (at $longer$ distances the mass induced by
Sun's gravitational field effectively decouples the scalar).

\subsubsection{Solar System observations}

Although the corrections that our theories introduce with respect
to GR for a spherically symmetric solution are suppressed by
powers of $(r/r_c)^4$, and this is a very small number for the Sun
within the Solar System, the precision with which the motion of
the celestial bodies of the Solar System is known makes it
conceivable that one could see the corrections induced by this
modification (similarly as happens in other long-distance
modifications of gravity as the DGP model \cite{Dvali:2002vf}).
For instance, in the case of the precession of the perihelion of
the planets, the anomalous shift ($\Delta \phi$) induced by a
small correction ($\delta V$) to Newton's potential ($V_N$) is
given in radians per revolution by (see $e.g.$
\cite{Dvali:2002vf}) \be \Delta \phi \simeq \pi r
\f{d}{dr}\left(r^2\f{d}{dr}\left(\f{\delta V}{rV_N}\right)\right),
\ee so the correction induced by our modification increases with
distance as \be \Delta \phi\simeq 16 \pi\left(\f{r}{r_c}\right)^4.
\ee From this expression we see that for the inner planets the
perihelion shift induced by our modification is very small, it
only becomes of the same order as the correction introduced by GR
(with respect to Newtonian gravity) for Jupiter. But for the
outer planets the contribution to the perihelion shift of the
modification can be dominant over the standard GR one\footnote{Not
only our
  correction increases with distance, but the GR one ($\Delta \phi \simeq 6\pi G_NM/r$) decreases for
  larger radius.}.
Unfortunately the only reliable data regarding the planetary
perihelion comes from the inner planets of the Solar System \cite{Pitjeva}, where
our correction is negligible. But the best
measured orbit is that of the Moon around the Earth by virtue of
the Lunar Laser Ranging (see $e.g.$ \cite{Williams:2004qb}). Using this the Earth-Moon distance is
known with a precision of centimetres. Applying the formula above, for the Moon our theory
predicts an anomalous shift of $\Delta \phi \sim 10^{-12}$, to be compared with the achieved accuracy of $2.4\times
10^{-11}$ \cite{Dvali:2002vf}. Although these numbers are not far from each other remember that we are just doing an order of magnitude estimation, since the parameter $\mu$ is related to $H_0$ only after choosing a particular function $f$. So although these theories suggest the possibility of surprises in high precision astrometrical measurements, the numbers that we obtained also expose the difficulties of ruling out these theories
using these effects, since for this we would need an improvement on the bounds of the anomalous precession of the Moon of at least two orders of magnitude.

\subsubsection{Cosmology and Astrophysics}

In this subsection we will comment on the possible comparisons
that one could make of the predictions of these theories to
astrophysical and cosmological observations (see for instance \cite{Aguirre:2003pg}), although any actual
fit to these data is beyond the scope of the current paper. In
this respect, once a particular function $f$ is chosen, one can
make unambiguous predictions for the rotation curves of spiral
galaxies with the mass-to-light ratio being the only free
parameter, since the normalisation of $\mu$ will be fixed by the
Hubble constant, and in fact this issue has been our main
motivation for proposing actions of the type (\ref{actionMOND}).
Our result for Newton's potential at short distances (\ref{sdsolII}) can be parameterised as: \be V(r)=\f{GM}{r}v(x)\,,\label{potential}\ee
where $x\equiv r/r_c$ and $v(x)\approx 1+\f{4}{3}x^4+\ldots$ for small $x$. The challenge now is to find a form for $f$ that yields a MOND-like phenomenology in the intermediate region. More specifically we need that $v(x)\sim - x\ln(x)$ for large $x$. The potential would then give rise to flat rotation curves that obey the Tully-Fisher law \cite{Tully:1977fu}. But
also other aspects of the observations of galactic dynamics can be
used to constrain a MOND-like modification of Newton's potential,
see $e.g.$  \cite{Zhao:2005zq}. And notice also that our theory
violates the strong equivalence principle, as expected for any
relativistic theory for MOND \cite{Milgrom:1983ca}, since locally
physics will intrinsically depend on the background gravitational
field. If we consider for instance a local system like an open
cluster, rotating in the background field of the Milky Way, the
solution (\ref{potential}) that we obtained for an isolated system is
not necessarily valid anymore. This will be the case if the background
curvature dominates the curvature induced by the local system,
similarly to the ``external field effect'' in MOND. Notice that in
order to establish the relevance of the external field effect in this
framework we have
to compare the induced external and internal curvatures, and not the
accelerations. This is
relevant for the case of the globular clusters orbiting the Milky Way,
where MOND (or DM)
evidence has been found where the internal acceleration is below
$a_0$, even if the external acceleration due to the
Milky Way is bigger than $a_0$ \cite{Scarpa:2003yw}. However, if one compares the values of
the curvature ($i.e.$, the invariant $Q$) one finds that the curvature is
dominated by the internal one, justifying the neglect of the external
field effect and the treatment of these
globular clusters as isolated objects, into the MOND regime. So the
globular cluster data of
\cite{Scarpa:2003yw} favors models, like the one presented in this paper, in
which the relevance of the external field effect is determined comparing $curvatures$,
and not accelerations.

At larger scales, where one can use the equivalence with a scalar-tensor theory more reliably, we can expect from continuity arguments that such a form for $f$ (if it exists) will enhance Newton's constant. 
One can then compare the theory against the observations of gravitational
lensing in clusters, the growth of large scale structure and the
fluctuations of the CMB. In fact, it has been pointed out that if GR was modified at large distances, an inconsistency between the allowed regions of parameter space would show up for (non-modified) Dark Energy models when comparing the bounds on these parameters obtained from CMB and large scale structure \cite{Ishak:2005zs}. This means that although some cosmological observables, like the expansion history of the Universe, can be indistinguishable in modified gravity and Dark Energy models, this degeneracy is broken when considering other cosmological observations and in particular the growth of large scale structure and the Integrated Sachs-Wolfe effect (ISW) have been shown to be good discriminators for models in which GR is modified \cite{Zhang:2005vt}.

Regarding the ISW effect, we would like to mention some characteristics of these theories that point to the possibility of getting a suppression of the low multipoles of the CMB with respect to a $\Lambda$CDM cosmology, a feature shown by the data that can not be easily accounted for in the $\Lambda$CDM model.
It has been recently pointed out that the fact that in the DGP model the effective Newton's constant increases at late times as the background curvature diminishes, causes a suppression of the ISW that brings the theory into better agreement with the CMB data than the $\Lambda$CDM model \cite{Sawicki:2005cc}. In our case we can expect an analogous effect, so that the effective Newton's constant for the cosmic perturbations depends on the background curvature in such a way that it increases at late times as the curvature diminishes. But despite this potentially good features it remains to be seen if one can get a good fit to the CMB data in the absence of Dark Matter in these models.

\section{Conclusions}

In this paper, motivated by the phenomenological success of MOND
fitting the rotation curves of spiral galaxies without requiring
Dark Matter, we have proposed a class of actions that modify
gravity below the characteristic acceleration scale required by
MOND: $a_0\sim H_0$. There are two effects in these theories that
are responsible for the infrared modification. First, there is an
extra scalar excitation of the spacetime metric besides the
massless graviton. The mass of this scalar field is of the order
of the Hubble scale in vacuum, but its mass depends crucially on
the background over which it propagates. This dependence is such
that this excitation becomes more massive as we approach any
source, and the extra degree of freedom decouples at short
distances in the spacetime of a spherically symmetric mass. This
feature makes this excitation to behave in a way that reminds of the chameleon field of
\cite{Khoury:2003aq}, but in our case this ``chameleon'' field is just a
component of the spacetime metric coupled to the curvature.
 But there is a second
effect in these theories: the Planck mass that controls the
coupling strength of the massless graviton also undergoes a
rescaling or ``running'' with the distance to the sources (or the
background curvature). This phenomenon, although a purely
classical one in our theory, is reminiscent of the quantum
renormalisation group running of couplings. So one might wonder if
actions of the type (\ref{actionMOND}) could be an effective
classical description of strong renormalisation effects in the
infrared that might appear in GR (see $e.g.$ \cite{Reuter:1996cp}
and references therein), as happens in QCD. In fact, corrections
depending on the logarithm of the renormalisation scale are
ubiquitous in quantum field theory, and it appears natural to
identify the renormalisation scale with a function of the
curvature if we want to build an effective classical action for
the spacetime metric that takes into account these quantum
effects. Indeed, we have seen that these models offer a phenomenology that
seems well suited to describe an infrared strongly coupled phase
of gravity: at high energies/curvatures we can use the GR action
or its linearisation as a good approximation, but when going to
low energies/curvatures we find a non-perturbative regime. At even
lower energies/curvatures perturbation theory is again applicable,
but the relevant theory is of scalar-tensor type in a de Sitter
space.

We would like also to emphasise that there are clearly many
modifications of the proposed class of actions that would offer a
similar phenomenology, such that gravity would be modified below a
characteristic acceleration scale of the order of the one required
in MOND. For instance if we consider the action \be S=\int
\!\!d^4x\sqrt{-g}\frac{1}{16\pi
  G_N}\left\{R-\mu^{2}\left({\rm
Log}[f]\right)^n\right\}\,,
\ee
with the same assumptions on $f$ that we did before we get that now
the critical acceleration $a_0$ also has a ``running'' with the mass as
\be
a_0=\f{G_NM}{r_c^2} \sim \mu\left({\rm Log}\left[\f{48a_0^3}{Q_0G_NM}\right]\right)^{(n-1)/2}.
\ee
To introduce some kind of scale dependence of $a_0$ could be interesting since MOND typically gives an overestimation of the amount of visible matter at cluster scales.

Since the expansions we have used break down for some
intermediate range of energies/distances one should still show that
the dynamics in this ``non-perturbative'' regime are consistent for
some choice of $f$ and that one recovers an acceptable matching between
the high energy/short distance and low energy/large distance
regimes. And then one should compute the predictions of these
theories in many different situations for which there are
experimental data to compare with before any of these models could
be considered a viable alternative to a $\Lambda$CDM cosmology.
This is a non-trivial task but it is worth undertaking it because
we have seen that there are reasons to believe that one might
explain many aspects of the cosmological and astrophysical
observations without introducing Dark Matter in this class of
theories. And, as we have seen, these theories also offer the
unique possibility of being tested not only through astrophysical
observations, but also through well-controlled laboratory experiments
where the outcome of such experiments is correlated with
parameters that can be determined by means of cosmological and
astrophysical measurements.

\section*{Acknowledgements}

We would like to thank S. Cole, A.C. Davis, D. Easson, G. Gibbons,
 A. Ibarra and J. Santiago
 for conversations and B. Solano for help with the plots. K.V.A. is supported by a postdoctoral grant of the Fund
for Scientific Research - Flanders (Belgium).

\end{document}